\documentclass[aps,pra,twocolumn,showpacs]{revtex4}
\usepackage{graphicx}
\usepackage{dcolumn}
\usepackage{bm}
\usepackage{epsfig}
\usepackage{amsmath}
\usepackage{amssymb}
\input{Qcircuit}

\newcommand{\one}{\mbox{$1 \hspace{-1.0mm}  {\bf l}$}}

\begin{document}

\title{Hybrid cluster state proposal for a quantum game}

\author{M. Paternostro, M. S. Tame, and M. S. Kim}

\affiliation{School of Mathematics and Physics, The Queen's University, Belfast, BT7 1NN, UK}
\date{\today}

\begin{abstract}
We propose an experimental implementation of a quantum game algorithm in a hybrid scheme combining the quantum circuit approach and the cluster state model.  An economical cluster configuration is suggested to embody a quantum version of the Prisoners' Dilemma. Our proposal is shown to be within the experimental state-of-art and can be realized with existing technology. The effects of relevant experimental imperfections are also carefully examined.
\end{abstract}

\pacs{03.67.-a, 02.50.Le, 03.67.Mn, 42.50.Dv}

\maketitle


\section{Introduction}

The study of the role played in a physical process by the quantumness of nature is a central issue for many branches of modern physics, leading to important and stimulating intellectual speculations~\cite{bell}. We have witnessed the possibilities offered by quantum mechanics as an exploitable resource which allows the accomplishment of tasks prohibitively difficult, if not impossible, in the classical domain. This point of view is particularly interesting in quantum information processing (QIP) where intrinsically quantum features such as entanglement are seen as valuable tools in applications like quantum cryptography, communication and computation. QIP can also be used in order to show the quantum behavior of nature and several studies along these lines have been performed. Non-classical states of electromagnetic fields~\cite{bowmeester}, entangled states of internal and external degrees of freedom of trapped ions~\cite{wineland} and the entanglement between atoms and light fields~\cite{monroe} are outstanding examples of this research. Recently, there has been considerable interest in showing the quantum behavior of macroscopic objects, paving the way toward a study of the boundary between classical and quantum physics~\cite{cantilever}. 

It is of the utmost importance to design simple experiments that provide evidence of the quantumness of nature. In this paper, we propose a readily available experimental protocol for quantum generalizations of classical games. We address how to implement a quantum version of the two-player Prisoners' Dilemma as an example of the distinctiveness of the intervention of quantum mechanics. Through quantum preparations and strategies~\cite{jensgiochi}, the players achieve goals which classically were impossible~\cite{johnssonPW}. Our proposal is based on the use of small multipartite entangled states, recently realized in all-optical setups~\cite{sara4fotoni,vlatko}. The choice of an optical scenario for the implementation of our proposal is incomparably suitable in studying the influences of quantum entanglement to the game. Indeed, even if the quantum game in~\cite{jensgiochi} has been implemented in a nuclear magnetic resonance (NMR) system~\cite{duNMR}, the density matrices of the highly mixed states involved in NMR can always be described as disentangled~\cite{braunstein}: the observation of ensemble-averaged pseudopure states renders the observation of the effects of the entanglement ambiguous~\cite{braunstein}. An all-optical implementation is not affected by this ambiguity. The multipartite entangled resource used in our proposal is given by cluster states~\cite{rb2} constructed through a double-pass scheme generating a four-photon entangled state via parametric down-conversion~\cite{vlatko} (the information is encoded in orthogonal photonic polarizations). 
\begin{figure}[b]
\vskip-0.6cm
\hspace*{1.2cm}{{\bf(a)}}\hspace*{3.3cm}{\bf(b)}\hspace*{1.4cm}{\bf (c)}\\
\hskip-4.5cm{$\cal P$}\hskip2.6cm${\cal M}$\\
\hskip-0.0cm
\begin{minipage}[l]{0.2\textwidth}
\centerline{\mbox{ \Qcircuit @C=0.85em @R=0.8em {
 & \gate{{\sf H}^{\cal P}_A}&\ctrl{1}&\gate{\sf H}&\gate{R_x^{-a}}&\ctrl{1}&
\gate{{\sf H}^{\cal M}_A}&\gate{\sf A}&\qw\\
&\gate{{\sf H}^{\cal P}_B}&\ctrl{-1}&\gate{\sf H}&\gate{R_x^{-b}}&\ctrl{-1}&
\gate{{\sf H}^{\cal M}_B}&\gate{\sf B}&\qw\gategroup{1}{8}{2}{9}{0.5em}{.}\gategroup{1}{2}{2}{3}{0.7em}{--}\gategroup{1}{6}{2}{7}{0.7em}{--}}}}
\end{minipage}
\hskip1.75cm
\begin{minipage}[c]{0.1\textwidth}
\centerline{\psfig{figure=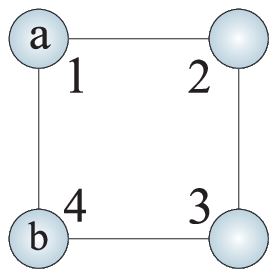,width=1.1cm,height=1.3cm}\hskip0.35cm\psfig{figure=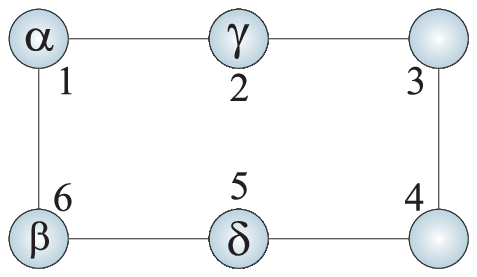,width=2.0cm,height=1.3cm}}
\end{minipage}
\caption{{\bf (a)}: Scheme of the quantum game. The input state is $\ket{c,c}_{AB}$, which evolves through ${\cal P}$ and ${\cal M}$ and the players' local strategies. Vertical lines denote control-phase gates, {\sf H}'s Hadamard gates and $R^{-(a,b)}_{x}$ single-qubit rotations around the $x$-axis. The dotted box is a utility stage. {\bf (b)} $\&$ {\bf (c)}: {\it Box} and {\it wafer} configuration for the sampling of the payoff. $\alpha,\beta,\gamma$ and $\delta$ are measurement angles.}
\label{formalscheme}
\end{figure}
Cluster states are a subclass of more general graph states and represent the quantum resource used in one-way computation, where one and two-qubit gates are simulated through a proper pattern of single-qubit measurements performed on a (sufficiently large) cluster state~\cite{rb2}.
The one-way model has triggered significant theoretical and experimental interest and schemes to improve the efficiency of linear optics computation have been proposed, based on this model~\cite{nielsenprl,nielsenbrownerudolph}. Cluster states naturally and economically simulate some operations which are central to our scheme~\cite{noiarchivio,noilungo}. We discuss how our proposal combines the standard quantum circuit model and cluster state based QIP, enhancing the one-way model. This {\it hybrid} model makes our scheme for a quantum game within the current state of the art. It is also one of the first immediately realizable protocols for quantum algorithms designed for small cluster configurations (for a two-qubit quantum search algorithm see Walther {\it et al.} in~\cite{vlatko}).

\section{The model} 

Let us consider the players $A$ and $B$ involved in the classical Prisoners' Dilemma, which is a non-zero sum game. The strategy space of each player is $S_{j}=\{c_{j},d_{j}\}$ ($j=A,B$). The game is non-cooperative and selfish, as the players aim to maximize their own payoff $\$_{j}(s)$, where $s$ is the strategy profile $s=(s_{A},s_{B})$ and $s_{j}\in{S_{j}}$ is the strategy chosen by player $j=A,B$~\cite{benjamin}. We have $\$_{A}(s)=0$ with $\$_{B}(s)=5$ (vice-versa) if the chosen profile is $s=(c_A,d_B)$ ($s=(d_A,c_B)$). When both players carry out the same strategy, the payoff is equally shared. They obtain the {\it cooperative payoff} (CP) $\$_{A,B}(s)=3$ if $s=(c_{A},c_{B})$, whereas they obtain the {\it equilibrium payoff} (EP) $\$_{A,B}(s)=1$ if $s=(d_{A},d_{B})$. A posteriori, $(d_{A},d_{B})$ is found to be a dominant profile. In fact, choosing $d_{j}$ and regardless of the strategy adopted by the adversary, player $j$ maximizes his payoff. The profile $(d_{A},d_{B})$ has the property that neither player can improve their payoff by a unilateral change of strategy, making it a {\it Nash equilibrium}~\cite{johnssonPW}. The rationality of the players and the non-cooperative nature of the game prevents $A$ and $B$ playing $(c_A,c_B)$ which is the {\it Pareto optimum}~\cite{johnssonPW}: no player can increase their payoff (which is the CP), by changing strategy, without reducing the payoff of the adversary. The Dilemma is in the dichotomy between the best choice for both and the highest payoff available individually. 

This Dilemma cannot be solved without some cooperativity. This is introduced in the quantum version of the game in~\cite{jensgiochi}, where the strategies which the players can use are embodied by a qubit [$\ket{c}=\begin{pmatrix}1\\0\end{pmatrix}$,\,$\ket{d}=\begin{pmatrix}0\\1\end{pmatrix}$]. Entangling stages ${\cal P}$ and ${\cal M}$ are introduced before and after the players perform their strategies. The strategy space is now $S_{j}\!=\!\{U_{j}(\theta_{j},\phi_{j})\vert\theta_{j}\!\in\![0,\pi],\!\phi_{j}\!\in\![0,\pi/2]\}$, where
\begin{equation}
\label{strategie}
U_{j}(\theta_{j},\phi_{j})=
\begin{pmatrix}
e^{-i\phi_{j}}\cos({\theta_{j}}/{2})&-\sin({\theta_{j}}/{2})\\
\sin({\theta_{j}}/{2})&e^{i\phi_{j}}\cos({\theta_{j}}/{2})
\end{pmatrix}
\end{equation}
and $c_{j}=U_{j}(0,0)$, $d_{j}=U_{j}(\pi,0)$. In~\cite{jensgiochi,benjamin}, the choice of $U_{A,B}$ and its consequences on the performances of the game are discussed. The entanglement gives $A$ and $B$ a degree of cooperativity. If their strategy profile $s=(U_{A},U_{B})$ is such that this cooperativity is preserved, a reconciliation between CP and EP can occur~\cite{commento2}. We stress that the procedure in~\cite{jensgiochi} is just one of the ways in which the game can be extended to the quantum realm. The choice in~\cite{jensgiochi} implies that the payoffs associated with $c_j$ and $d_j$ will be the classical values and a Pareto optimal point is sought from the additional strategies provided by the quantum strategic space.

In general, the less restrictive constraint imposed on the quantum version of a protocol is that it reproduces the classical process, in the proper limiting case. Here, this means that the description of the Prisoners' Dilemma when $\cal{P}$ and ${\cal M}$ are removed must match the classical one. At the same time, we look for a generalized game where the payoffs are affected by the entanglement so as to provide a Pareto optimal point lying within the strategy space of a separable game.

The structure of the entangling steps is dictated by the interaction naturally realized by the setup considered. In our case, ${\cal P}$ and ${\cal M}$ must be related to the two-qubit gates simulated by a particular cluster configuration. In this respect, it is important to notice that a simple two-qubit cluster state results in the effective simulation of a controlled $\pi$-phase gate ($\sf CP_{\pi}$)~\cite{noilungo}. This is the key advantage with respect to non-cluster based standard quantum circuit schemes. Our proposal is able to naturally embody nearly the entire quantum steps ${\cal P}$ and ${\cal M}$, which otherwise, have to be implemented by two independent two-qubit operations. This is because networking these operations to obtain the scheme in Fig.~\ref{formalscheme} {\bf (a)} is in general a difficult task. The use of a cluster state in our proposal, represents a major advantage in this respect. In addition, in the same two-qubit cluster, the measurement of a qubit in the basis $\{\ket{\pm}^{a}=\ket{c}\pm{e}^{ia}\ket{d}\}$ simulates the application of $R^{-a}_{x}{\sf H}$ on a logical qubit, where $R^{-a}_{x}$ is a rotation by an angle $-a$ around the $x$-axis of the Bloch sphere and ${\sf H}$ the Hadamard gate. The full quantum circuit we propose is shown in Fig.~\ref{formalscheme} {\bf (a)}, with the part prior to the dotted box being simulated by the cluster in Fig.~\ref{formalscheme} {\bf (b)}. The state corresponding to this {\it box cluster}, introduced by Walther {\it et al} in~\cite{vlatko}, can be put into the form
$\ket{box}=({1}/{4})[\ket{0}_{1}+\ket{1}_{1}(\sigma_{z,2}\!\otimes\!\sigma_{z,4})]({\sf H}_{2}\otimes{\sf H}_{4})\ket{ghz}_{234}$, where $\ket{ghz}$ is a GHZ state. We now exploit the naturally simulated $\sf CP_{\pi}$ gate and ${\sf H}^{\cal P}_{j}$'s (implicit in the preparation of a cluster state~\cite{rb2}) to obtain ${\cal P}={\sf CP_{\pi}}({\sf H}^{\cal P}_{A}\otimes{\sf H}^{\cal P}_{B})$. The ${\sf H}^{\cal P}_{j}$'s allow us to generate a maximally entangled strategic state and to combine superpositions of orthogonal strategies and entanglement~\cite{commento}. Despite the conceptual equivalence of ${\cal P}^{\dag}$ and ${\cal M}=({\sf H}^{\cal M}_{A}\otimes{\sf H}^{\cal M}_{B}){\sf CP_{\pi}}$, it is worth differentiating them as ${\sf H}^{\cal M}_{j}$'s are simulated in the box cluster by measuring qubits $2$ and $3$ in the $\sigma_{x}$ eigenbasis.
\begin{figure} [b]
\vskip-0.6cm
\hspace*{0.3cm}{{\bf(a)}}\hspace*{4.3cm}{\bf(b)}
\psfig{figure=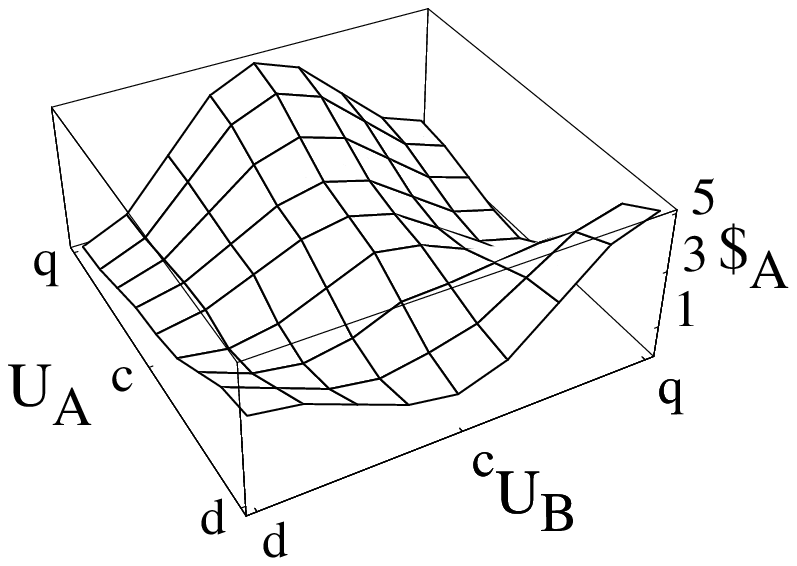,width=4.5cm,height=3.7cm}\psfig{figure=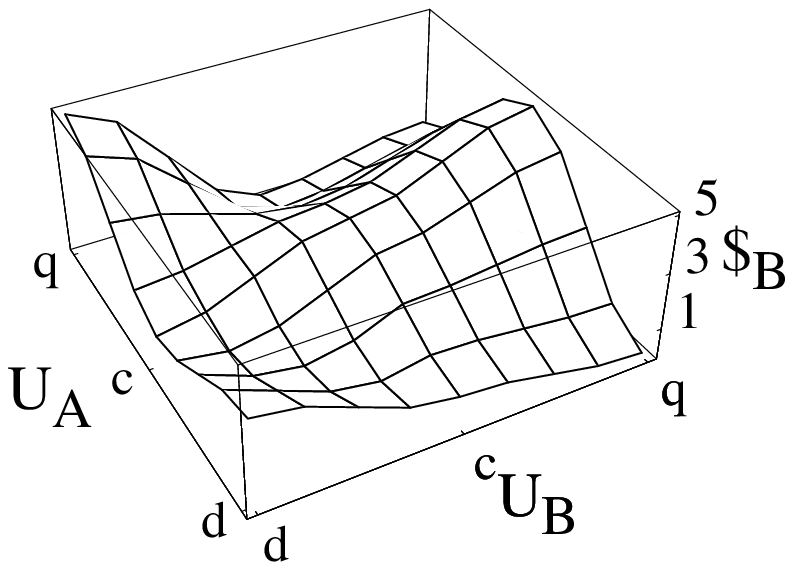,width=4.5cm,height=3.7cm}
\caption{{\bf (a)}: $\$_A$ vs. the strategies $U_{A,B}$. 
 {\bf (b)}: $\$_B$ vs. $U_{A,B}$. In both the panels, the parameterization is $U_{j}=U_{j}(p\pi,0)$ for $p\in[0,1]$ and $U_{j}=U_{j}(0,-p\pi/2)$ for $p\in[-1,0)$ ($j=A,B$). Here, $d_{j}$ corresponds to $p=1$, $c_{j}$ to $p=0$ and $q_{j}$ to $p=-1$.}  
\label{payoff}
\end{figure}

Quantitatively, we need to calculate the expression 
\begin{equation}
\label{amplitudes}
P_{\xi\chi}=\vert\phantom{}_{AB}\langle{\xi,\chi}\vert{\cal M}(U_{A}\otimes{U}_{B}){\cal P}\ket{c,c}_{AB}\vert^2\hskip0.2cm(\xi,\chi=c,d),
\end{equation}
which gives the probability that the evolved strategy profile, after the operations by the players, is $s=(\xi_{A},\chi_{B})$. With Eq.~(\ref{amplitudes}) it is easy to evaluate $\$_{A}(U_A,U_B)=3P_{cc}+P_{dd}+5P_{dc}$ and $\$_{B}(U_A,U_B)=3P_{cc}+P_{dd}+5P_{cd}$.
The results are shown in Figs.~\ref{payoff} {\bf (a)} and {\bf (b)}. The strategic sector $[c_{j},d_{j}]$ ($[q_{j},c_{j})$) corresponds to $\phi_{j}=0$ ($\theta_{j}=0$) with $\theta_{j}\in[0,\pi]$ ($\phi_{j}\in[0,\pi/2]$).
We take this parameterization as it reveals the relevant features of the game. 

From Fig.~\ref{payoff} {\bf (a)}, we see that for $B$ choosing $d_{B}$ or $q_B$, the best strategy by $A$ is $d_{A}$ with payoffs $\$_{A}(d_{A},d_{B})=3$ or $\$_{A}(d_{A},q_{B})=5$ respectively. 
Analogous considerations can be made {\it mutatis mutandis} about $\$_{B}(U_{A},U_{B})$ (Fig.~\ref{payoff} {\bf (b)}). It can be seen that the profile $(d_{A},d_{B})$ is the only Nash equilibrium. The players' payoff for this profile is exactly the CP, which shows that $(d_{A},d_{B})$ is Pareto optimal. This result is quantum mechanical, as the payoff corresponding to $(d_{A},d_{B})$, in a game without ${\cal P}$ and ${\cal M}$, is EP. Indeed, in the separable quantum game resulting from the removal of ${\sf CP}_{\pi}$'s in ${\cal P}$ and ${\cal M}$ (keeping ${\sf H}^{{\cal P},{\cal M}}_{A,B}$), no reconciliation is attained. Moreover, later we show that the Pareto optimality cannot be attained by using classical correlations shared between the players of the game, suggesting that the entanglement provided to the players favors the reconciliation of the Dilemma. While the procedure in~\cite{jensgiochi} introduces a new strategy profile which is a Nash equilibrium and achieves CP, in our scheme the equilibrium strategy is the same as in the non-entangled game. The entanglement renders $(d_{A},d_{B})$ the profile preserving the cooperativity introduced by ${\cal P}$.

Parts of the game are naturally implemented by a box cluster, but the strategies $U_{A,B}$ must be simulated by an appropriate measurement pattern. As shown in Figs.~\ref{formalscheme} {\bf (a)} and {\bf (b)}, by measuring qubits $1$ and $4$ we can simulate just a rotation around the $x$ axis of the single-qubit Bloch sphere~\cite{noilungo}. Thus, we need more freedom for the players to perform their strategies. For this task, we exploit the fact that ${\sf H}^{\cal M}_{A,B}$ and ${\sf CP}_{\pi}$ belong to the Clifford group. We consider the operations ${\sf A,B}\in\{\sigma_{x,y,z},R^{\mu}_{x}\}$ in the dotted box of Fig.~\ref{formalscheme} {\bf (a)}, which can be {\it imported} to the dashed section of the circuit. They are seen as operations on the qubits $2,3$ of the box cluster applied before their final measurement. Together with $R^{-a,-b}_{x}$ simulated by the measurement of $1$ and $4$, these enlarge the strategy space of the players. In Table~\ref{tabella1}, we show the measurement angles $a$ and $b$ and the corresponding ${\sf A}$, $\sf B$. Only two measurement bases are needed and $\one$ or $\sigma_{x}$ must be imported before the measurements are performed. 

We remark that the use of local operations on the logical output qubits of a cluster is inherent to the one-way model~\cite{rb2}. The randomness of the measurement outcomes affects a gate simulation which has to be corrected by local decoding operators. Here, we are implicitly assuming the postselection of those events corresponding to the projection of qubits $1$ and $4$ onto $\ket{+}^{a,b}_{1,4}$. In this case, the decoding operators are $\one_{2,3}$. ${\sf A}$ and ${\sf B}$ may be seen as decoding operators selected not by the measurement outcomes but by the task to perform. The hybrid nature of our approach should be clear: we cannot rely just on the measurement-based gate simulations because we need additional rotations of the logical output qubits. Here, ${\sf A}$ and ${\sf B}$ can be easily realized in the all-optical setups in~\cite{vlatko}. Indeed, by exploiting that ${\sf H}^{\cal M}_{j}\sigma_{x,j}=\sigma_{z,j}{\sf H}^{\cal M}_{j}$, the players only need to apply $\sigma_{z}$ to the output qubits, which is possible via phase shifters, 
before they are measured in the $\sigma_{x}$ eigenbasis. \begin{figure} [b]
\vskip-0.6cm
{\bf (a)}\hskip3.8cm{\bf (b)}
\centerline{\psfig{figure=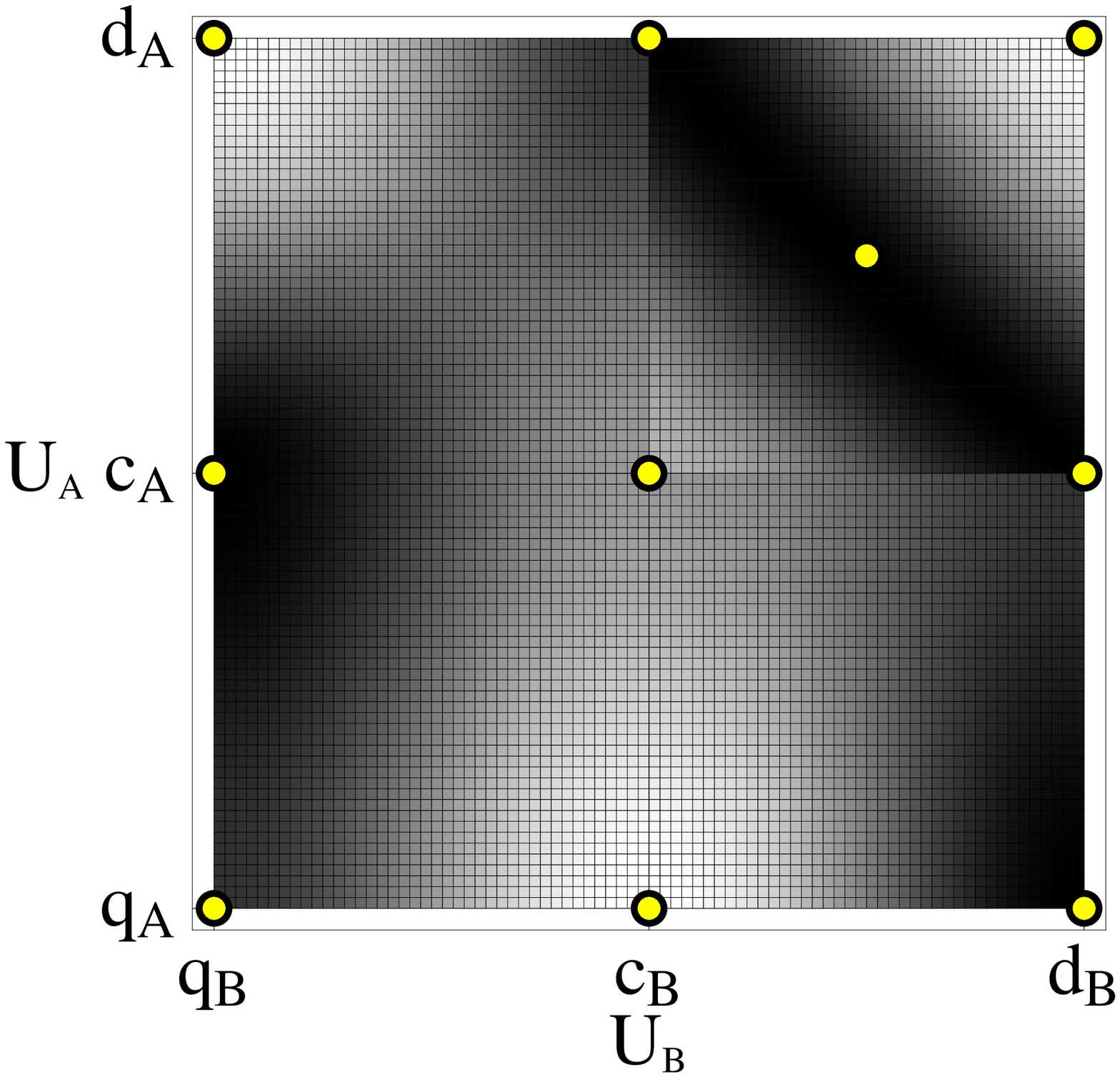,width=4.0cm,height=3.5cm}\hskip0.5cm\psfig{figure=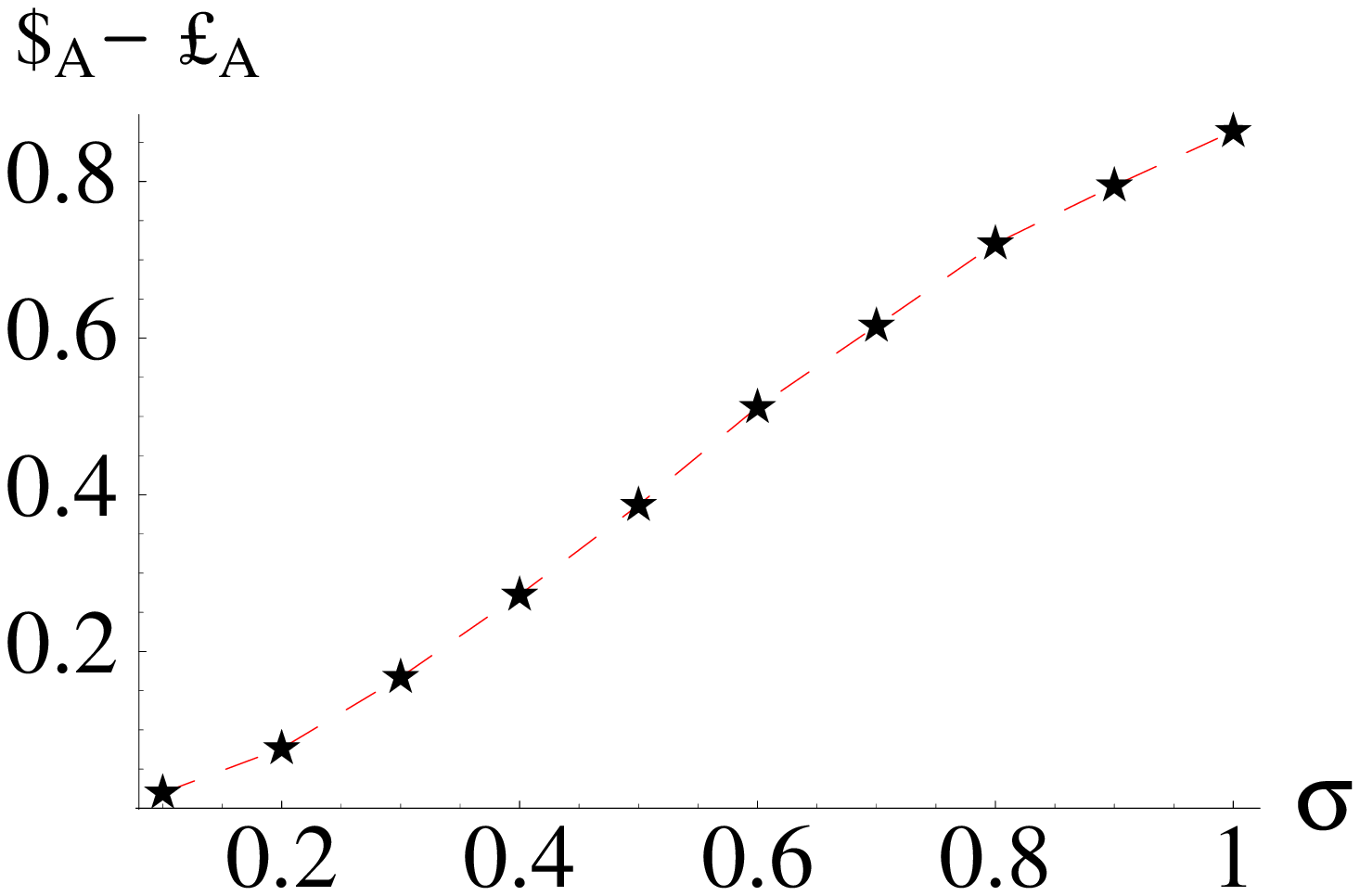,width=4.5cm,height=3.2cm}}
\caption{{\bf (a)}: Density plot of $\$_{A}$. The brighter the plot, the higher the payoff. Each bright dot shows a strategy in Table~\ref{tabella1}. The central dot in the top-right quadrant corresponds to the profile $(m_A,m_B)$. {\bf (b)}: Differences between $\$_{A}(d_{A},d_{B})$, the ideal payoff, and \pounds$_{A}(d_{A},d_{B})$, the average with imperfections, plotted against the standard deviation $\sigma$ of the normal distributions attached to the rotation angles.}  
\label{imperfezioni}
\end{figure}
\begin{table} [t]
\begin{ruledtabular}
\begin{tabular}{|l|c|c|c|c|c|}\hline
Strategy&$-a\hskip0.05cm-b$&${\sf A}\hskip0.4cm{\sf B}$&Strategy&${-a\hskip0.05cm-b}$&${\sf A}\hskip0.6cm{\sf B}$\\ \hline\hline
${c_{A}\hskip0.5cm{c}_{B}}$&$0\hskip0.5cm{0}$&$\hskip-0.1cm\one\hskip0.5cm\one$& ${q_{A}\hskip0.5cm{d}_{B}}$&$0\hskip0.5cm{\pi}$&${i}\sigma_{x}\hskip0.2cm{i}\sigma_{x}$\\ \hline
${c_{A}\hskip0.5cm{q}_{B}}$&$0\hskip0.5cm{0}$&$\hskip0.1cm\one\hskip0.4cm{i}\sigma_{x}$&${d_{A}\hskip0.5cm{c}_{B}}$&$\pi\hskip0.5cm{0}$&$\hskip-0.1cm{i}\sigma_{x}\hskip0.4cm\one$\\ \hline
${c_{A}\hskip0.5cm{d}_{B}}$&$0\hskip0.5cm{\pi}$&$\hskip0.1cm\one\hskip0.4cm{i}\sigma_{x}$ &${d_{A}\hskip0.5cm{q}_{B}}$&$\pi\hskip0.5cm{0}$&${i}\sigma_{x}\hskip0.2cm{i}\sigma_{x}$\\ \hline
${q_{A}\hskip0.5cm{c}_{B}}$&$0\hskip0.5cm{0}$&$\hskip0.1cm{i}\sigma_{x}\hskip0.3cm\one\hskip0.3cm$&${d_{A}\hskip0.5cm{d}_{B}}$&$\pi\hskip0.5cm{\pi}$&${i}\sigma_{x}\hskip0.2cm{i}\sigma_{x}$\\ \hline
${q_{A}\hskip0.5cm{q}_{B}}$&$0\hskip0.5cm{0}$&$\hskip0.3cm{i}\sigma_{x}\hskip0.2cm{i}\sigma_{x}\hskip0.3cm$&${m_{A}\hskip0.4cm{m}_{B}}$&$\pi\hskip0.5cm{\pi}$&$\one\hskip0.6cm\one$\\ \hline
\end{tabular}
\end{ruledtabular}
\caption{Rotation angles and imported operations for the strategies in the quantum game with $m_{A,B}=\frac{1}{\sqrt{2}}(\one+i\sigma^{(A,B)}_{y})$.\label{tabella1}}
\end{table}
In this way, all the strategies in Table~\ref{tabella1} can be attained, which is sufficient to experimentally study the Pareto-optimality of the Nash equilibrium point (Fig.~\ref{imperfezioni} {\bf (a)} graphically shows these strategies). However, this does not exhaust all our possibilities. For instance, the entire {quadrant} $[q_{A},c_{A}]\times[q_{B},c_{B}]$ can be sampled simply by taking $a=b=0$, importing ${\sf A}=R^{\mu}_{x},\,{\sf B}=R^{\nu}_{x}$ and scanning the angles $\mu,\nu$. Single-qubit manipulations through linear-optical elements just prior to the detection stage make our scheme feasible~\cite{vlatko}. 

However, we cannot sample the entire payoff $\$_{A,B}$ with the box cluster because it is not possible to obtain $R^{\theta_{j}}_{y}=R^{-\pi/2}_{x}R^{\theta_{j}}_{z}R^{\pi/2}_{x}\equiv{U}_{j}({\theta_{j},0})$. For a complete tomography of $\$_{A,B}$, the price to pay is the use of a larger number of qubits. Indeed, using the concatenation technique~\cite{noilungo} and an analysis similar to the one relative to the box cluster, it can be seen that the {\it wafer} configuration in Fig.~\ref{formalscheme} {\bf (c)} can fully embody this quantum game. In addition, the rotation $R_{y}^\theta$ can be realized by choosing $\alpha=\beta=\pi/2,\,\gamma=\theta_{A},\,\delta=\theta_{B}$ and importing ${\sf A}={\sf B}=R^{\pi/2}_{x}$, which correspond to a phase shift $R^{\pi/2}_{z}$ applied to $3$ and $4$ before measuring in the $\sigma_{x}$ eigenbasis. The wafer configuration is realized by gluing two four-photon entangled states~\cite{sara4fotoni} using the technique suggested in~\cite{nielsenbrownerudolph} and realized by Zhang {\it et al.} in~\cite{vlatko}.

\section{Effects of imperfections}

We now address the effect of realistic imperfections in this game. Non-idealities come from errors introduced at the measurements. The waveplates in front of the photodetectors used to measure the state of the cluster qubits may introduce unwanted rotations of a polarization state, leading to wrong measurement bases. In addition, imperfections at the down-conversion stage in generating a box cluster provide mixed entangled states to the players. Both these sources of error can be formally considered by the replacement $\theta_{j}\rightarrow\theta_{j}+\epsilon_{j}$ in Eq.~(\ref{strategie}) (analogously for $\phi_{j}$) and averaging the payoffs over appropriate probability distributions, with standard deviation $\sigma_{j}$, attached to $\epsilon_{j}$'s. This randomness results in a corrupted mixed entangled resource~\cite{imoto} whose degree of entanglement diminishes if $\sigma$ is increased. In Fig.~\ref{imperfezioni} {\bf (b)} we show the differences between the ideal (Pareto optimal) $\$_{A}(d_{A},d_{B})$ and the average payoff \pounds$_{A}(d_{A},d_{B})$ obtained when $\epsilon_{j}$'s are normally distributed around $0$. The result is not affected by fluctuations in $\phi_{j}$ as $d_{j}$ does not depend on this parameter. At $\sigma\simeq0.9$ the degree of entanglement (quantified by the measure based on the Peres-Horodecki criterion~\cite{npt}) is $\lesssim{0.01}$. The larger the fluctuations allowed for $\epsilon_{j}$, the larger the deviation of the corresponding payoff from the behaviors in Fig.~\ref{payoff}. The effect of classical correlations can also be studied by considering the mixed initial state $\otimes^{B}_{j=A}[(1-x)\ket{c}_{j}\!\bra{c}+x\ket{d}_{j}\!\bra{d}]$ ($x\in[0,{0.5}]$) to enter ${\cal P}$, resulting in a non-ideal entangled mixed resource which $A$ and $B$ use to play the described game. For $x\ge0.29$, this mixed state is separable so that $A$ and $B$ only share classical correlations. In this case, it is easy to find that CP$>\$^{x\ge{0.29}}_{A,B}(d_{A},d_{B})>$EP. Also, no other Pareto optimal points arise as a result of different strategy profiles. This provides an operative way to study how the Pareto optimality is lost when corrupted resources and imperfect measurements are used in our scheme. It represents a useful practical and theoretical tool in studying the performances of the quantum game.


\section{Remarks} 

We have proposed an experimental implementation of the quantum Prisoners' Dilemma through an economical and experimentally realizable cluster state configuration. At the same time, we have shown that the cluster model can be complemented by simple rotations of the logical output qubits to add freedom to the gate simulation, building a hybrid model which can be realized with existing technology. This allows for an immediate experimental investigation of the role of entanglement in the search for a Pareto optimal Nash equilibrium point in a system exhibiting quantum correlations. 
\newline
\acknowledgments

MSK thanks P. Walther and M. Aspelmeyer for discussions. This work has been supported by the UK EPSRC, KRF (2003-070-C00024), The Leverhulme Trust (ECF/40157) and DEL.


\begin{thebibliography}{99}

\bibitem{bell} J. S. Bell, {\sl Speakable and Unspeakable in Quantum Mechanics} (Cambridge University Press, Cambridge, 1987).


\bibitem{bowmeester}  
D. Bowmeester {\it et al.}, {\sl Nature }{(London)} {\bf 390}, 575 (1997).

\bibitem{wineland} C. Monroe, D. M. Meekhof, B. E. King, and D. J. Wineland, {\sl Science} {\bf 272}, 1131 (1996).

\bibitem{monroe} J.M. Raimond, M. Brune, and S. Haroche, {\sl Rev. Mod. Phys.} {\bf 73}, 565 (2001); B. B. Blinov, D. L. Moehring, L.-M. Duan, and C. Monroe, {\sl  Nature}{ (London)} {\bf 428}, 153 (2004).

\bibitem{cantilever} K. Schwab, E. A. Henriksen, J. M. Worlock, and M. L. Roukes, {\sl Nature}{ (London)} {\bf 404}, 974 (2000); C. Brukner, V. Vedral, and A. Zeilinger, {\sl quant-ph/0410138}.

\bibitem{jensgiochi} J. Eisert, M. Wilkens and M. Lewenstein, {\sl Phys. Rev. Lett.} {\bf 83}, 3077 (1999); {\sl Phys. Rev. Lett.} {\bf 87}, 069802 (2001).

\bibitem{johnssonPW} C. F. Lee and N. F. Johnson, {\sl Phys. World}, {\bf 15}, 25 (2002).

\bibitem{vlatko} P. Walther {\it et al.}, {\sl Nature} {\bf 434}, 196 (2005); A.-N. Zhang, {\it et al.} {\sl quant-ph/0501036}.

\bibitem{sara4fotoni} P. Walther {\it et al.}, {\sl Nature} (London), {\bf 429}, 158 (2004).

\bibitem{duNMR} J. Du {\it et al.}, {\sl Phys. Rev. Lett.} {\bf 88}, 137902 (2002).

\bibitem{braunstein} S. L. Braunstein {\it et al.}, {\sl Phys. Rev. Lett.} {\bf 83}, 1054 (1999).

\bibitem{rb2} R. Raussendorf and H. J. Briegel, {\sl Phys. Rev. Lett.} {\bf 86}, 5188 (2001). 

\bibitem{nielsenbrownerudolph} D. E. Browne and T. Rudolph, {\sl quant-ph/0405157}.

\bibitem{nielsenprl} M. A. Nielsen, {\sl Phys. Rev. Lett.} {\bf 93}, 040503 (2004).

\bibitem{noiarchivio} M. S. Tame, M. Paternostro, M. S. Kim, and V. Vedral, {\sl quant-ph/0412156}.

\bibitem{noilungo} M. S. Tame, M. Paternostro, M. S. Kim, and V. Vedral, {\sl Phys. Rev. A} {\bf 72}, 012319 (2005).  

\bibitem{benjamin} S. C. Benjamin, and P. M. Hayden, {\sl Phys. Rev. A} {\bf 64}, 030301 (2001); {\sl Phys. Rev. Lett.} {\bf 87}, 069801 (2001).

\bibitem{commento2} We do not consider issues related to the fact that the players are provided with the aid of an entangled state [see S. J. van Enk and R. Pike, {\sl Phys. Rev. A} {\bf 66}, 024306 (2002)] as this does not imply the possibility for them to coordinate their strategies.

\bibitem{commento} D. A. Meyer, {\sl Phys. Rev. Lett.} {\bf 82}, 1052 (1999).

\bibitem{imoto} N. F. Johnson, {\sl Phys. Rev. A} {\bf 63}(R), 020302 (2001); S.K. Ozdemir, J. Shimamura, and N. Imoto, {\sl quant-ph/0402038}.

\bibitem{npt}  A. Peres, {\sl Phys. Rev. Lett.} {\bf 77}, 1413 (1996); M. Horodecki, P. Horodecki, and R. Horodecki, {\sl Phys. Lett. A} {\bf 223}, 1 (1996); J. Lee, M. S. Kim, Y. J. Park, and S. Lee, {\sl J. Mod. Opt.} {\bf 47}, 2151 (2000).

\end{thebibliography}
\end{document}